\documentclass[a4paper]{article}   %% LaTeX 2e (preferred)

\usepackage{amsmath}
\usepackage{mathptmx}						%Times Font for latin and Symbol font for greek.
\usepackage[english]{babel}
\usepackage[T1]{fontenc}
\usepackage{hyperref} 
\usepackage[pdftex]{graphicx}
\usepackage{cite}					
\usepackage{ifpdf}
\usepackage{color}
\usepackage{inputenc}
\usepackage{authblk}

\begin{document} 

\title{Experimental investigation of relaxation oscillations resonance in mode-locked Fabry-Perot semiconductor lasers}

\author[1]{V.~Roncin*,~J.~Poëtte,~J-F.~Hayau, ~P.~Besnard,~J-C.~Simon}
\author[2]{F.~Van~Dijk,~A.~Shen ~and ~G-H.~Duan}

\affil[1]{CNRS UMR 6082 Foton, Enssat 6 rue de Kerampont, F-22300 Lannion Cedex, France}
\affil[2]{Alcatel-Thales III-V Lab, Route Départementale 128, F-91767 Palaiseau Cedex, France }
\date{}

\thanks{Corresponding author information \\
E-mail: \textit{vincent.roncin@enssat.fr}, \\
Telephone: +33 (0)2 96 46 91 50}

\maketitle

\begin {abstract}
We propose in this communication an experimental study of the relaxation oscillations behavior in mode-locked lasers. The semiconductor self-pulsating laser diode is composed by two gain sections, without saturable absorber. It is made of bulk structure and designed for optical telecommunication applications. This specific device allows two different regimes of optical modulation: the first one corresponds to the resonance of the relaxation oscillations and the second one, to the mode-locking regime at FSR value. This singular behavior leads us to characterize the self-pulsations which are coexisting in the laser and to describe two regimes of output modulation: the first one appears thanks to the resonance of the oscillation relaxation and the other one corresponds to the FSR of the Fabry-Perot laser at 40 GHz.
\end {abstract}

%%%%%%%%%%%%%%%%%%%%%%%%%%%%%%%%%%%%%%%%%%%%%%%%%%%%%%%%%%%%%
\section{INTRODUCTION}

Recent advances in semiconductor lasers sources propose new devices allowing to generate high-frequency optical pulsed sources compatible with optical communications requirements. This type of lasers is based on passive mode-locking at the frequency imposed by the laser geometry\cite{phamJQE98} and particularly imposed by the length of the cavity. The telecommunication applications such as high bit-rate optical transmitter demand very stable and high spectral purity mode-locked lasers based on self-pulsation process\cite{bobbertEL04}\,. Other specific applications such as all-optical regeneration\cite{simonECOC98} impose specific optical injection properties for the all-optical clock recovery stage at high frequencies above 40\,GHz\cite{sartoriusEL98,RoncinPTL2007}\,. This all-optical telecommunication context require high performances without electrical modulation neither external feedback. One of the most studied self-pulsating lasers is designed using multi-sections allowing the combination of optical properties such as phase control and wavelength tuneability\cite{DuanPTL1995}\,. Recently, it has been proposed a simple Fabry-Perot semiconductor laser offering high-speed optical performances characterized by short pulses generation compatible with telecommunication applications\cite{GossetAPL2006,LavigneJLT2007}\,. The principle of the self-pulsating process is partially described in the literature\cite{RadziunasJQE2000} but without a clear explanation of the mode-locking starting. 
Furthermore, classical references on passive mode-locking process in lasers\cite{SiegmanAPL1969} and particularly in semiconductor lasers\cite{HausJQE1975} describe it in one hand thanks to the saturable absorption allowing a passive phase selection of the optical modes, and on the other hand, thanks to an internal modulation or perturbation at the harmonic frequency of the Free Spectral range (FSR)\cite{HausSTQE2000,JonesJQE1995}\,. 
Thus, the origin of the self-pulsation in multimode semiconductor lasers without saturable absorber is usually attributed to nonlinear mechanisms of interaction similar to wave-mixing between the optical modes in the gain region.          
We propose in this paper a pure experimental study of the self-pulsating semiconductor lasers starting and the behavior of the relaxation oscillations. We described firstly the self-pulsating laser provided in the frame-work of the French research program ROTOR and particularly their dynamic properties at 42\,GHz. Spectral and time domain analysis of the output optical modulation is observed at the relaxation frequency and at the FSR frequency. Then, we propose a study of the coexistence of both type of modulations induced by two different process observed simultaneously in the laser.

%%%%%%%%%%%%%%%%%%%%%%%%%%%%%%%%%%%%%%%%%%%%%%%%%%%%%%%%%%%%%
\section{SELF-PULSATING LASER CHARACTERISATION}

\begin{figure} [!h]
	 \centering
	 \includegraphics[height=6cm]{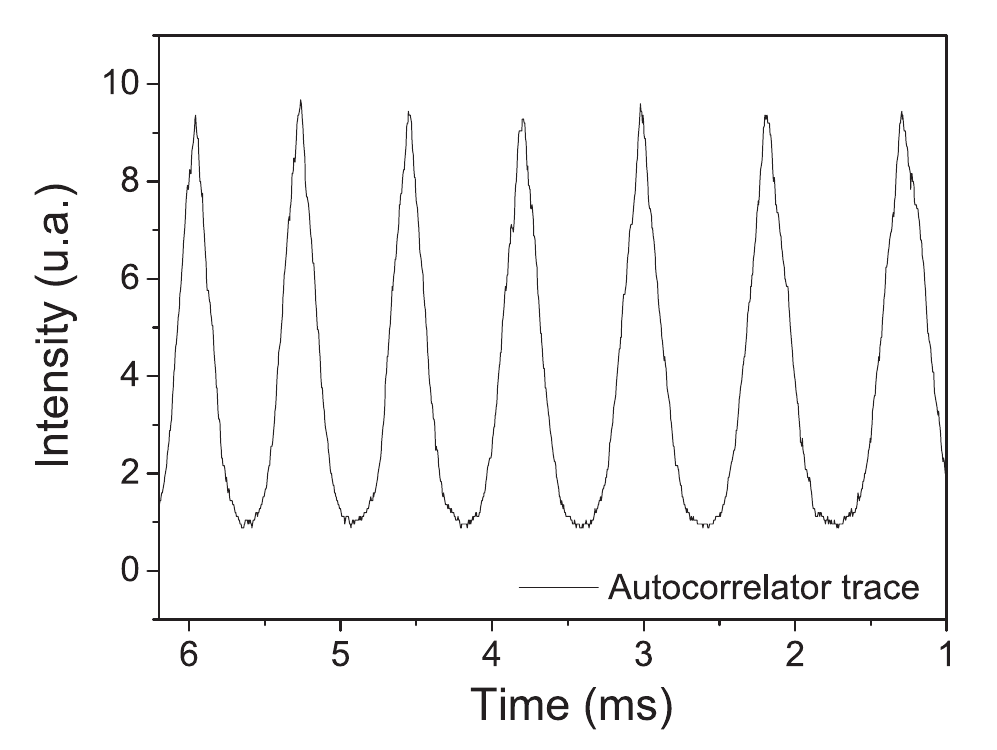}
	 \label{fig1} 
	 \caption{Autocorrelation trace when the laser is self-pulsating at 40 GHz: the deconvulated pulsewidth is 5.5 ps.}
   \end{figure}

\begin{figure} [!h]
	 \centering
	 \includegraphics[height=6cm]{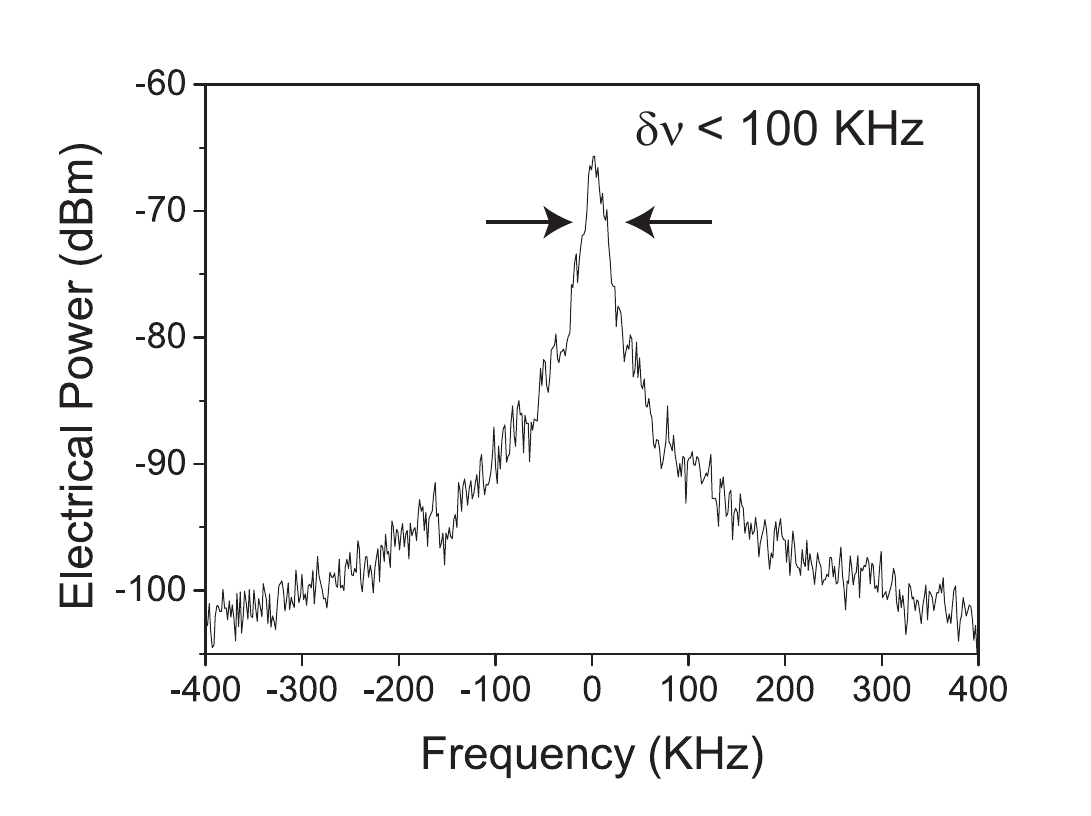}
	 \label{fig2}
	 \caption{Electrical bandwidth of the self-pulsation. The narrowest modulation is about 100\,KHz at a frequency of 42.2\,GHz for the couple of current $\{100,40\}$.}
   \end{figure}

\begin{figure}[!h]
	 \centering
   \includegraphics[height=6cm]{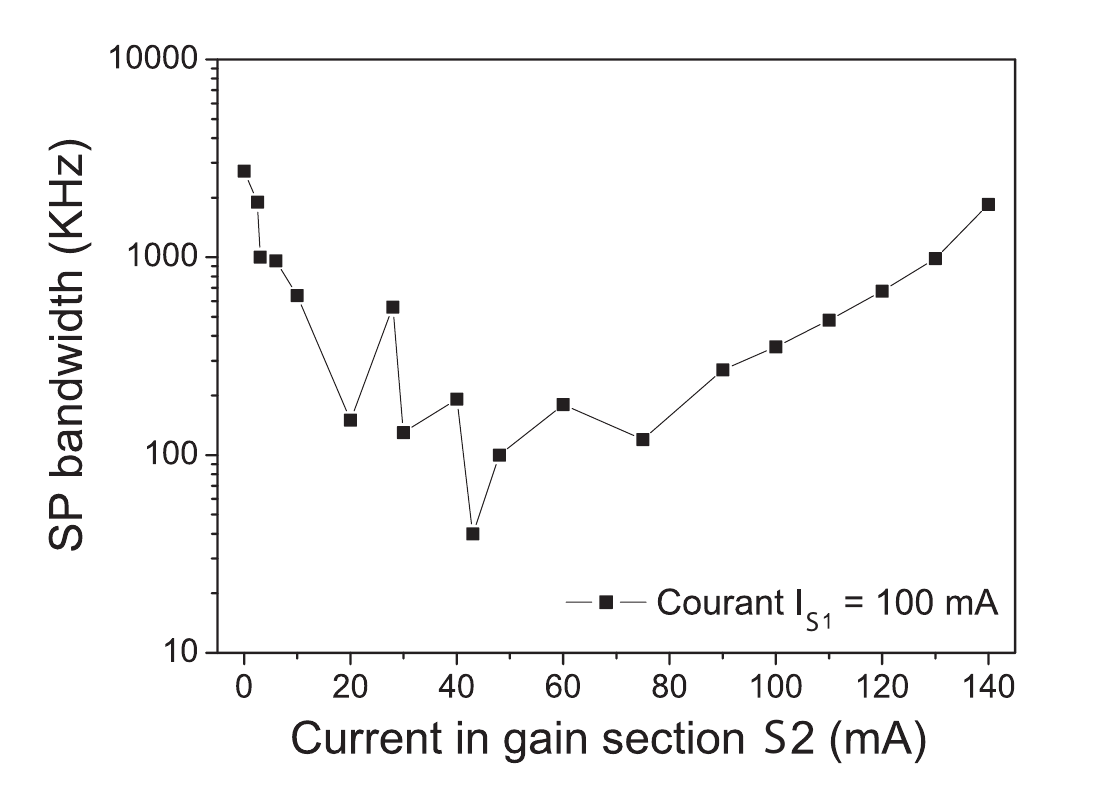}
   \label{fig3} 
   \caption{Evolution of the modulation bandwidth when the current of the secondary gain section \emph{S\,2} is varing and the current in \emph{S\,1} is fixed. The minimum linewidth is less than 100\,KHz for 40\,mA in \emph{S\,2}.}
   \end{figure}

The laser studied in the following experimental report is a Fabry-Perot Bulk semiconductor laser with a length of about 970 $\mu$m emitting over a bandwidth of 10\,nm at 3\,dB centered at 1.6\,$\mu$m \cite{VandijkECOC2006}\,. The ridge waveguide, which is 1.3\,$\mu$m wide over two gain sections, is buried with bulk p-doped InP. The two sections are electrically isolated from the others by ion implantation of the inter-electrode regions. The principal section designated at \emph{S\,1} corresponds to about 90\% of the laser length. The secondary section is 10\% of the laser length and will be called in the following \emph{S\,2}. Firstly we realized the basic characterisation of the laser output power with a laser threshold $I_{th}^L$ corresponding to a couple of current values equal to 18 mA in the \emph{S\,1} and 0~mA in the \emph{S\,2}: $\{I_1,I_2\}$ = $\{18,0\}$. 
The self-pulsating threshold $I_{th}^{SP}$ has been measured at about 4-times $I_{th}^L$ corresponding to a value of 73\,mA. We observed that it is necessary to polarize the \emph{S\,2} in order to induce self-pulsations. Nevertheless, the current in the \emph{S\,2} does not modify the laser threshold when it is higher than 4~mA. Let's stress that the \emph{S\,2} does not behave as a saturable absorber because it is definitely a gain section, identical to the \emph{S\,1}. Figure~\ref{fig1} shows the classical autocorrelation trace of the optical pulse-train with a deconvoluted value of 5.5\,ps for the pulsewidth and a repetition rate around 25\,ps corresponding to a frequency of about 40\,GHz. This value is close to the value of the FSR of the Fabry-Perot cavity. This temporal trace was obtained without any optical bandwidth filtering of the laser spectrum. Moreover we have been able to obtain shorter pulsewidth by using narrowband optical filters. The filter allows the chirp influence to be reduced and the pulsewidth to be close to the Fourier Transform limit. This point implies that the mode-locking of the optical modes present a chirp according to the wideness of the spectrum.
\\
In this paper, we propose a characterization of the self-pulsation process. This measurement is obtained thanks to an electrical spectrum analyzer. The relaxation frequency evolution is directly measured in the band of the electrical spectrum analyzer corresponding to a frequency limitation at 8\,GHz. Besides, the self-pulsation modulation at 42.2\,GHz is too high for our equipment. So we resolved this by using a specific frequency converter which transposes the high frequency in the the band of the analyzer.  
The narrowest electrical linewidth obtained with our device is presented in figure~\ref{fig2}, when the laser is self-pulsating. The width of about 100\,KHz corresponds to a very high spectral purity of the optical modulation at 42.2\,GHz. This result is obtained for the couple of current values in the \emph{S\,1} and \emph{S\,2} $\{100,40\}$.
Figure~\ref{fig3} gives the evolution of the linewidth at a fixed current in the principal gain section \emph{S\,1} when a variable current is injected in the gain section \emph{S\,2}. It shows clearly that the linewidth is varying within 2 orders of magnitude and that is very sensitive to the current value.

%-------------
%------------- 

%-------------
%-------------

%-------------
%-------------

The high threshold observed for the self-pulsation as compared to the low laser threshold allows the laser relaxation frequency resonance to be studied, in the range of current for \emph{S\,1}: [$I_{th}^L$,$I_{th}^{SP}$].
In the following part using both the electrical spectrum analyzer and a real-time oscilloscope the resonance of the relaxation oscillations, which is leading to a deep optical modulation, are analyzed by using an electrical spectrum analyzer and a high-bandwidth oscilloscope. 

%%%%%%%%%%%%%%%%%%%%%%%%%%%%%%%%%%%%%%%%%%%%%%%%%%%%%%%%%%%%%
\section{HIGH RESONANCE OF THE RELAXATION OSCILLATIONS}

%%-----------------------------------------------------------
\subsection{Frequency and Time domain observation} 

\begin{figure} [!h]
   \centering
   \includegraphics[height=19 cm]{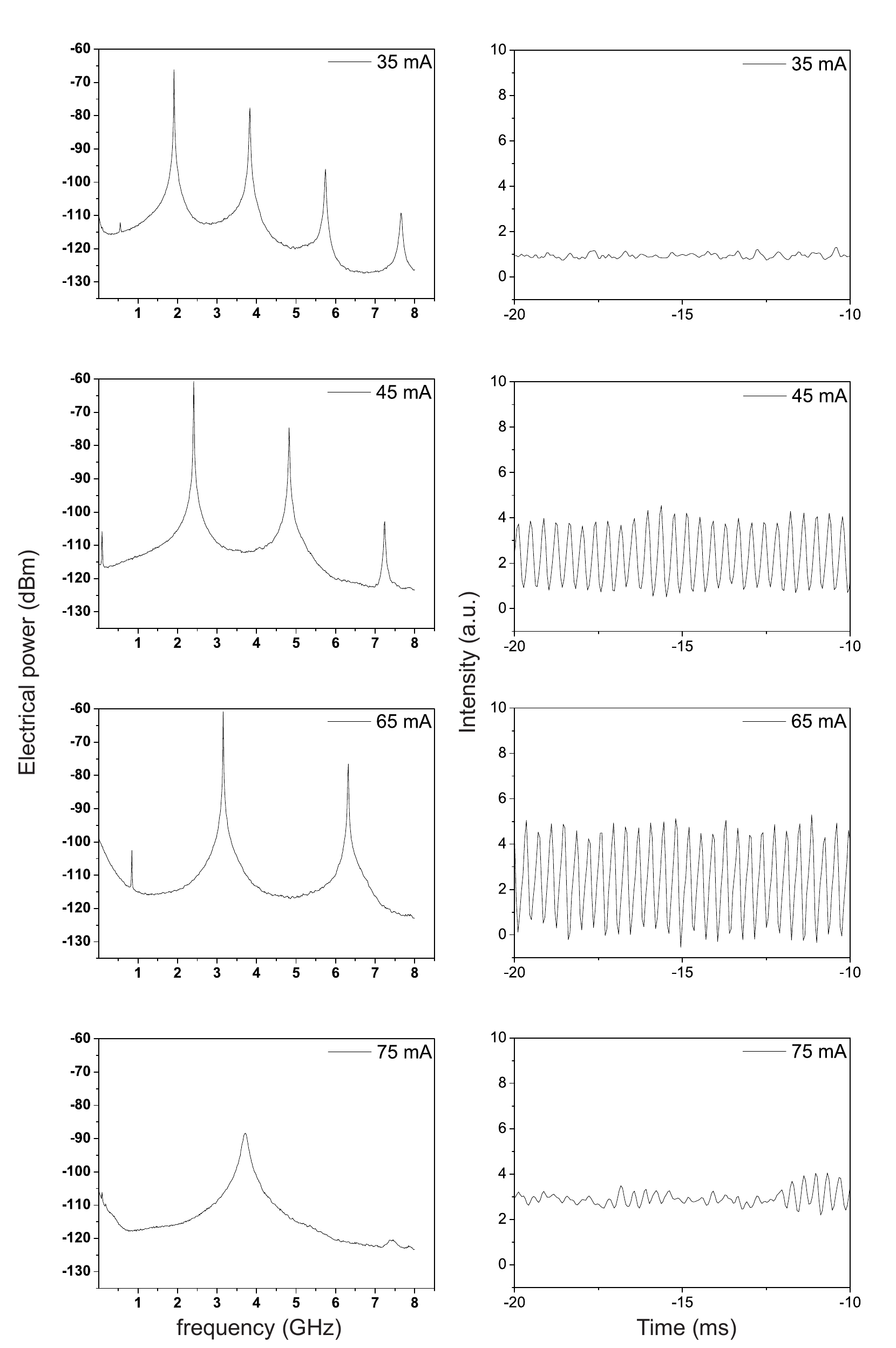}
   \label{fig4} 
   \caption{Frequency and time domain experimental analysis of the resonance of the relaxation oscillations and its harmonics in the self-pulsating Fabry-Perot laser. The increase of the current in the \emph{S\,1} is changing the magnitude of the modulation. The current in the secondary gain \emph{S\,2} is fixed to 2\,mA.}
   \end{figure}

This section is devoted to the analysis in the time and spectral domain of the resonance induced by the carrier relaxation. Former works have already proposed numerical studies and experimental observations of oscillation resonance due to carrier relaxation in semiconductor lasers\cite{VahalaAPL1983,OvadiaPTL1992} or due to optical feedback in coupled cavity\cite{TagerJQE1994}. This phenomena leads to an output self-pulsation allowing to generate an optical modulation enhanced by a resonance effect.
In our laser, an important optical modulation at the characteristic relaxation frequency is observed both in the time domain using a real time oscilloscope with a wide bandwidth up to 4\,GHz and simultaneously in the frequency domain using an electrical spectrum analyzer.
The observations of the phenomena are plotted in figure~\ref{fig4}. The curves on the left hand figure presents the peak of resonance at $f_{relax}$ and the harmonics. The curves on the right hand figure present the corresponding modulation observed in the time domain. We show the evolution of the phenomena by increasing the current in \emph{S\,1} for a fixed current in \emph{S\,2} equal to few milliamps. The modulation observed in the time domain appears for ${I_{S1}\,\approx\,40~mA}$, under the self-pulsation threshold $I_{th}^{SP}$, and disappears as soon as ${I_{S1}}$ is reaching the $I_{th}^{SP}$ value.
Besides, the relaxation frequency is moving with the injected current in \emph{S\,1} in accordance with the behavior proper to semiconductor bulk lasers\cite{PoetteSPIE2007}\,.

%-------------
%-------------

%%-----------------------------------------------------------
\subsection{Optical domain observation} 

\begin{figure}[!h]
   \centering
   \includegraphics[height=6cm]{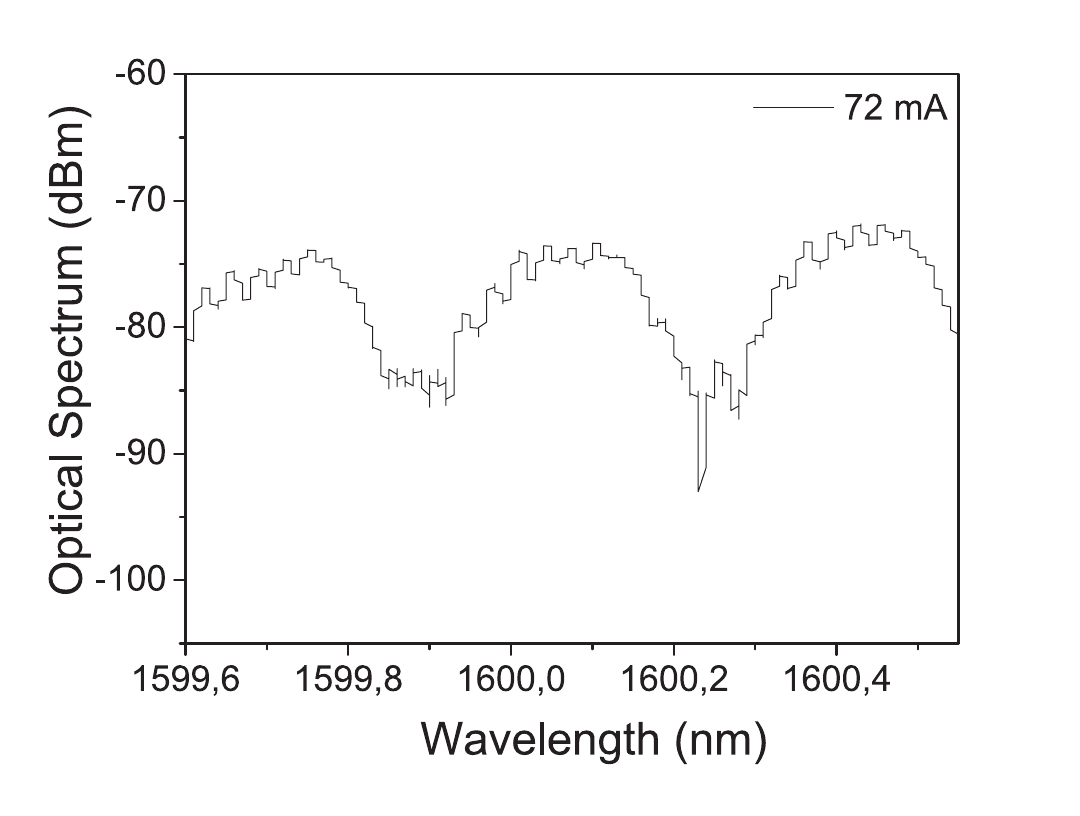}
	 \label{fig5} 
	 \caption{Optical spectrum observation of the Fabry-Perot modes under the self-pulsation threshold $I_{th}^{SP}$. $I_1$\,=\,72\,mA and $I_2$\,=\,2\,mA. The modulation is observed on the spectrum thanks to the 10\,pm of resolution for the optical spectrum analyzer}
   \end{figure}

The characterization of the resonance is made thanks to an optical spectrum analysis with 10\,pm of resolution and high sensitivity equipment. Figure~\ref{fig5} shows the optical spectrum of the laser under the self-pulsating threshold at the maximum of optical power. A modulation of the Fabry-Perot modes is clearly observed. The experimental measurement of the modulation gives a value close to the relaxation frequency. The limited accuracy of the equipment does not allow a precise measurement of the frequency in the optical spectral domain. On the other hand, it is possible to observe that the over-modulation seems to be disymmetrical with regard to the Fabry-Perot peaks.
In figure~\ref{fig6}, the optical spectrum of the laser is given, at the same wavelength, above the self-pulsating threshold. Regarding to the symmetry of the modulation lines with the Fabry-Perot modes we deduce that they are in phase with the modulation at the relaxation frequency. 

%-------------
%------------- 

These results allow us to understand the phase correlation process at the FSR frequency. Such as it has been proposed in the laser theory of mode-locking without saturable absorption\cite{VahalaJQE1983}, the gain modulation at a harmonic frequency of the FSR leads to lock in phase the optical modes. Thus, the quality of the mode-locking is enhanced thanks to the nonlinearities in the laser\cite{RenaudierPTL2005} and the estimation of the phase correlation is deduced from phase noise measurement\cite{TsuchidaOL98}.
In the following section we propose an additional result allowing to synthesize the coexisting of the two modulations in the lasers and to describe the transfer of power between relaxation oscillation and self-pulsation at the FSR.

%%%%%%%%%%%%%%%%%%%%%%%%%%%%%%%%%%%%%%%%%%%%%%%%%%%%%%%%%%%%%
\section{THE SELF-PULSATION AT THE FSR} 

\begin{figure}[!h]
	 \centering
   \includegraphics[height=7cm]{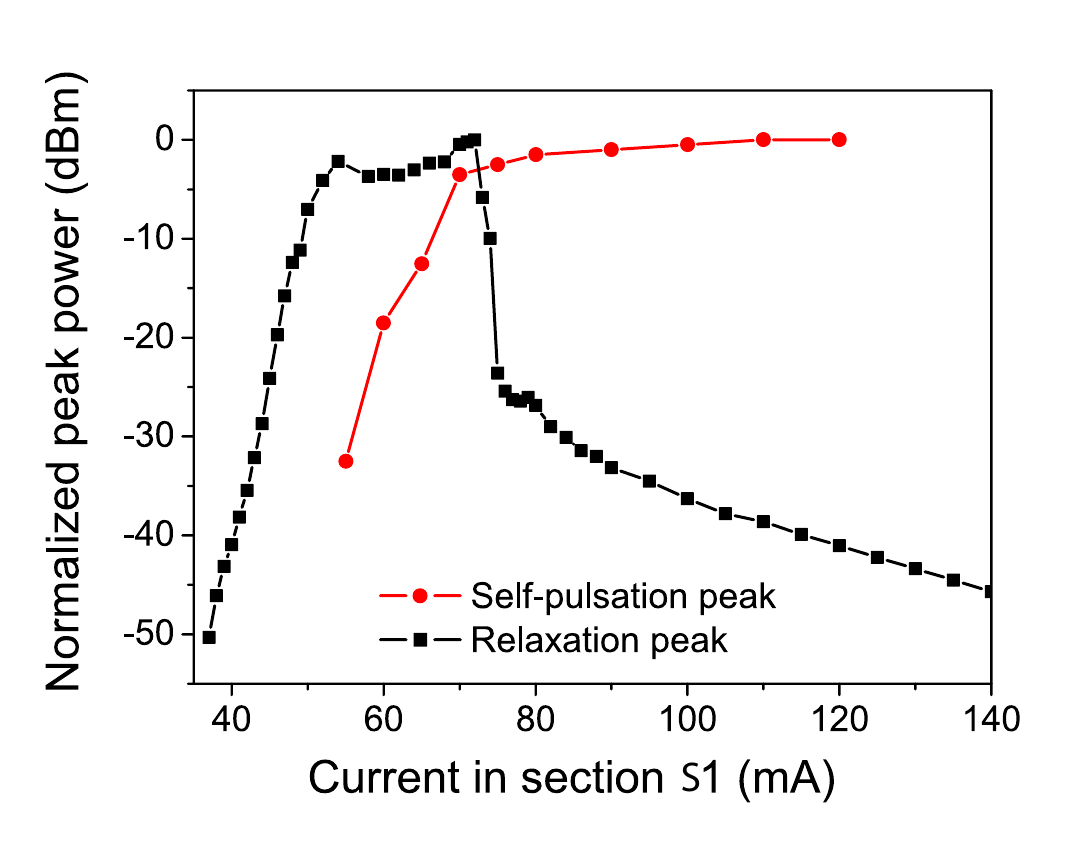}
   \label{fig7}
   \caption{Experimental measurement of the electrical power evolutions of the coexisting optical modulations. The self-pulsation modulation at the FSR frequency (circles) and the modulation induced at the resonance of the relaxation oscillations (squares).}
   \end{figure}

We propose hereafter to synthesize the experimental observations by plotting the power evolution of the relaxation oscillations resonance and comparing it with the evolution of the self-pulsation modulation. We change the peak powers of the modulations by varying the drive current in \emph{S\,1} with a fixed current of 2\,mA in \emph{S\,2}.
The results are presented in figure~\ref{fig7}.
Two different regimes are clearly observed: the first one corresponds to a current $I_1$ under $I_{th}^{SP}$ equal to 73 mA and the second one to the above $I_{th}^{SP}$. These two regimes are both characterized by the presence of a strong optical modulation: in the first one, at the relaxation frequency and in the second one, at the FSR frequency.  
In the first range of current, the resonance of the relaxation oscillations is increasing until a saturation which is then stopped in close to the Self-pulsation threshold $I_{th}^{SP}$.
Thanks to this comparison, we show that when the self-pulsation at the FSR reaches a sufficient value, the modulation at the relaxation frequency decreases. Thus, when the self-pulsation regime starts, the relaxation oscillations disappear. Nevertheless this singular  optical modulation transfer does not allow further hypothesis about a possible correlation between the two self-pulsation regimes.

%-------------
%-------------  

\clearpage
%%%%%%%%%%%%%%%%%%%%%%%%%%%%%%%%%%%%%%%%%%%%%%%%%%%%%%%%%%%%%
\section{CONCLUSION} 

We reported an experimental investigation of the self-pulsating process in a Fabry-Perot laser made of bulk semiconductor structure. The high resonance of the relaxation oscillations under the mode-locking threshold and the birth of the self-pulsation regime at the FSR frequency are experimentally observed. The particularly high self-pulsation threshold at the FSR frequency, equal to 4-times the laser threshold, allow us a fine comparison between the two coexisting regimes of output modulations. We particularly propose results  on an hypothetical transfer of modulation from relaxation oscillation to self-pulsation.

%%%%%%%%%%%%%%%%%%%%%%%%%%%%%%%%%%%%%%%%%%%%%%%%%%%%%%%%%%%%%
\section{acknowledgments}     %>>>> equivalent to \section*{ACKNOWLEDGMENTS}       
 
This work has been supported by the "R\'egion Bretagne" and the "Minist{\'e}re de la recherche" in the framework of the research programs DisTO and ROTOR. Foton is also supported by the European Found for Regional Development (FEDER). The authors want to specially thanks all the projects partners for fruitful discussions.

%%%%%%%%%%%%%%%%%%%%%%%%%%%%%%%%%%%%%%%%%%%%%%%%%%%%%%%%%%%%%

\bibliographystyle{IEEEtran}   %>>>> bibliography data in mode_lock_starting.bib
\bibliography{IEEEabrv,mode_lock_starting}   %>>>> makes bibtex use spiebib.bst (bibliographic style)

\end{document}